\documentclass[aip,jcp,amsmath,amssymb,reprint]{revtex4-1}

\usepackage{graphicx}
\usepackage{dcolumn}
\usepackage{bm}
\usepackage{color}
\usepackage{hyperref}
\usepackage{multirow}
\usepackage[normalem]{ulem}

\newcolumntype{d}[1]{D{.}{.}{#1}}

\usepackage[utf8]{inputenc}
\usepackage[T1]{fontenc}
\usepackage{mathptmx}

\newcommand{\half}{\frac{1}{2}}

\begin{document}

\preprint{AIP/123-QED}

\title{Determination of fundamental properties of nitrogen from first principles. \\ I. Atomic polarizabilities and long-range dispersion coefficients}

\author{Micha\l\ Przybytek}
\author{Jakub Lang}
\author{Micha\l\ Lesiuk}
\email{e-mail: m.lesiuk@uw.edu.pl}
\affiliation{\sl University of Warsaw, Faculty of Chemistry, Pasteura 1, 02-093 Warsaw, Poland}
\date{\today}

\begin{abstract}
This work is the first in a series of papers in which we perform theoretical calculations of various fundamental properties of nitrogen relevant for gas thermometry experiments. In this part we focus on the properties of nitrogen atom, namely dynamic polarizabilities and dispersion coefficients that describe interaction between two nitrogen atoms at large internuclear separations. These quantities are calculated using a composite scheme based on coupled cluster and full configuration interaction methods and large Gaussian basis sets optimized specifically for the purposes of this work. The dispersion coefficients, $C_n$ with $n=6,8,10$, are obtained using Casimir--Polder formulas by numerical integration over dipole, quadrupole, and octuple polarizabilities for imaginary frequencies represented analytically by Pad\'e approximants. Special attention is paid to careful error control and uncertainty estimation of the calculated quantities.
\end{abstract}

\maketitle

\section{Introduction}
\label{sec:intro}

Measurements of pressure of a gas are of fundamental importance in many branches of science and industry alike. Current pressure standards are based on experimental methods such as dielectric constant gas thermometry (DCGT) \cite{gaiser15,gaiser17,gaiser18,gaiser20,gaiser21,gaiser22} or refractive index gas thermometry (RIGT). \cite{jousten17,gao17,rourke19,ripa21,rourke21} Taking the second method as an example, pressure of the working gas, $p$, is inferred from the direct measurement of its refractive index, $n$. To establish a link between these two quantities, one combines the generalized Lorentz-Lorenz formula \cite{lorentz80,lorenz80} and the equation of state of a given gas. In the low pressure regime, this leads to the expression
\begin{align}
\label{ll}
    p = \frac{kT}{2\pi [\alpha(\omega) + \chi(\omega)]}(n-1),
\end{align}
where $T$ is the temperature, $k$ is the Boltzmann constant (having a fixed numerical value since 2019 according to the new definition of the SI units), and $\alpha(\omega)$ and $\chi(\omega)$ are the polarizability and magnetic susceptibility, respectively, of an atom/molecule of the working gas. The latter two quantities are dependent on the frequency of light, $\omega$, at which the measurement is taken. Similar expressions are used in the DCGT method, the difference being that the dielectric constant of the gas is measured instead of the refractive index. For higher pressures, it becomes important to extend the formula~(\ref{ll}) by including higher-order terms in the gas density. The coefficients multiplying these terms are expressed through the so-called density and dielectric virial coefficients. The complete working formulas of the DCGT and RIGT methods that include such terms can be found, for example, in Refs.~\onlinecite{gaiser15,rourke19}.

The choice of the working gas is one of the most important considerations in gas thermometry experiments. The current primary pressure standard is based on helium gas. The advantage of this medium, in addition to its exceptional stability and chemical inertness, is the fact that all necessary fundamental properties of helium can be calculated by theoretical means with sufficiently high accuracy. \cite{pachucki00,lach04,piszczatowski15,puchalski16,puchalski20,garberoglio20,czachorowski20,garberoglio21,lang23a,lang23b,puchalski23,garberoglio24,binosi24} This makes the approach based on helium truly primary and independent of previous measurements. Unfortunately, the setup based on helium has also certain drawbacks, the most important being probably the sensitivity to impurities. In the microwave regime, the most common impurity, water vapor, has polarizability larger roughly by a factor of $20$ in comparison with helium. Therefore, even a small admixture of such impurities into the working gas has a large influence on the results. This necessitates the usage of ultrapure gases, which increases the cost and difficulties in handling the experimental setup. To address this issue, the use of heavier noble gases, primarily neon and argon, has been proposed recently and has already been demonstrated in the RIGT experiments, see Ref.~\onlinecite{rourke21} for an extended overview of the topic. These advances have been accompanied by the determination of relevant properties of neon and argon gases using theoretical methods. \cite{wiebke12,lesiuk20,garberoglio21b,hellmann21,hellmann22,lesiuk23,lesiuk24a,lesiuk24b} However, as the cost of theoretical calculations rapidly increases with the number of electrons, the precision of the results is in most cases significantly lower in comparison with helium.

Another possible choice of the working gas is to abandon the atomic gases altogether and consider molecular gases which are sufficiently stable and chemically nonreactive within the necessary range of pressure (up to several MPa) and temperature (below $2000\,$K). However, the use of an arbitrary molecule for this purpose would be problematic, because most molecules possess a permanent electric dipole moment. This property makes Eq.~(\ref{ll}) no longer valid and compromises the main idea of gas thermometry experiments due to decreased sensitivity to the pressure of the gas. All above requirements drastically restrict the catalog of candidate molecules. Taking a glance at the periodic table, one quickly finds that the simplest candidates for such applications are H$_2$, N$_2$ or O$_2$. The nitrogen is particularly attractive due to its stability, chemical inertness as well as broadly availability in academia, metrology institutes, industry, etc.

Despite the broad use of nitrogen, there is a surprising lack of data available for this system that are relevant to the DCGT and RIGT experiments. The most important quantities in this context are the static and dynamic polarizabilities and the magnetic susceptibility, see Eq.~(\ref{ll}). This work is the first in a series of papers with the goal of filling this gap and providing reliable theoretical data, including rigorous uncertainty estimates, for the properties in question. Unlike for helium, it is impossible to determine all necessary properties of nitrogen purely from theoretical calculations with sufficient accuracy from the point of view of DCGT and RIGT methods. Therefore, to assemble the required data, one uses a patchwork of experimental and theoretical results that complement each other, similarly as for neon and argon. Quantities which are difficult to measure or do not need to be known extremely accurately are usually obtained by theoretical means. An illustrative example of such a procedure is given in the last paper of the series.

When molecular gases are used in thermometry experiments, an additional complication arises in comparison with noble gases which makes such hybrid experimental-theoretical approach particularly attractive. Note that the polarizability and magnetic susceptibility of an isolated atom do not depend on temperature. This is no longer true for a diatomic molecule, where all quantities of interest carry a dependence on the bond length and thus on the temperature at which the measurement is taken. This makes direct measurement of $\alpha(\omega,T)$ over the whole range of frequencies and temperatures an even more daunting proposition. Consequently, their theoretical calculation becomes a viable alternative to direct measurement. Similar considerations apply to the magnetic susceptibility of nitrogen.

This work is the first part of a series of papers devoted to the theoretical determination of the aforementioned fundamental properties of nitrogen. The outline of this series of papers is as follows. In this work (Paper~I), we determine the static and dynamic polarizability of the nitrogen atom and use these results to calculate the dispersion coefficients $C_n$, which describe the interaction of two nitrogen atoms at large internuclear distances. Although these quantities are not used directly in gas thermometry experiments, they will be key ingredients in subsequent papers. Moreover, knowledge of atomic polarizabilities and dispersion coefficients is valuable from the point of view of other applications such as in density functional theory for adding the missing long-range dispersion effects \cite{grimme10,grimme11} or in development of force fields used in molecular mechanics simulations. \cite{karplus02,walters18} In Paper~II, we present a new theoretical potential energy curve for the interaction of two nitrogen atoms with rigorous uncertainty estimates. This enables us to find with controlled accuracy the temperature dependence of any property of nitrogen molecule. Finally, in Paper~III we calculate the dynamic polarizability and magnetic susceptibility for two interacting nitrogen atoms at relevant internuclear distances. After proper account of both electronic and nuclear contributions, we determine the dynamic polarizability of nitrogen, $\alpha(\omega,T)$, with explicit frequency and temperature dependence that can be used directly in gas thermometry experiments.

\section{Theory}
\label{sec:theory}

In this work, we consider two fundamental quantities of the nitrogen atom. The first is the $2^l$-pole dynamic polarizability at an arbitrary frequency $\omega$ defined as \cite{stone13}
\begin{align}
\label{alphaw1}
    \alpha_l(\omega) = 2\sum_{n\neq0} \frac{\omega_n}{\omega_n^2-\omega^2}\,
    |\langle n| Q_{l0} | 0\rangle|^2,
\end{align}
or without the sum over states
\begin{align}
\label{alphaw2}
\begin{split}
    \alpha_l(\omega) 
    &= \langle0|Q_{l0} \big( H - E_0 + \omega \big)^{-1} Q_{l0} |0\rangle \\
    &+ \langle0|Q_{l0} \big( H - E_0 - \omega \big)^{-1} Q_{l0} |0\rangle,
\end{split}
\end{align}
where $H$ is the electronic Hamiltonian of the system, $|0\rangle$ is the electronic ground state of the atom with energy $E_0$, $|n\rangle$ denotes the $n$-th excited electronic state with energy $E_n$, $\omega_n=E_n-E_0$ is the electronic excitation energy, and $Q_{l0}$ is the $2^l$-pole moment operator defined by the general expression
\begin{align}
\label{defQlm}
  Q_{lm}=-\sqrt{\frac{4\pi}{2l+1}}\sum_i r_i^l\,Y_{lm}(\hat{r}_i),
\end{align}
where the summation runs over all electrons in the atom, the radial and angular coordinates of the $i$-th electron are denoted as $r_i$ and $\hat{r}_i$, respectively, and $Y_{lm}(\hat{r})$ are the standard spherical harmonics.

The second family of quantities considered in this work are the dispersion coefficients $C_n$, with $n=6, 8, 10$, which describe the interaction between nitrogen atoms at large internuclear distances. They are defined using the Casimir--Polder formulas as \cite{casimir46,casimir48}
\begin{align}
\label{c6cp}
    &C_6 = \frac{3}{\pi} \int_0^\infty d\omega\;\big[\alpha_1(i\omega)\big]^2,
\end{align}
\begin{align}
\label{c8cp}
    &C_8 = \frac{15}{\pi} \int_0^\infty d\omega\;\alpha_1(i\omega)\,\alpha_2(i\omega),
\end{align}
and
\begin{align}
\label{c10cp}
\begin{split}
    C_{10} &= \frac{28}{\pi} \int_0^\infty d\omega\;\alpha_1(i\omega)\,\alpha_3(i\omega) \\
    &+ \frac{35}{\pi} \int_0^\infty d\omega\;\big[\alpha_2(i\omega)\big]^2.
\end{split}
\end{align}

Note that in the case of dynamic polarizabilities, we are interested only in experimentally relevant real frequencies below the first resonance frequency. However, to evaluate the dispersion coefficients, we also need the polarizabilities at purely imaginary frequencies. The former quantities can be evaluated directly using linear response theory from numerous wavefunction models, as described in more detail in Sec.~\ref{sec:numeric}. While the same approach is in principle possible also for imaginary frequencies, the formalism is more complicated in this case, and direct determination of $\alpha_l(i\omega)$ is not implemented for most methods used in this work. Therefore, we require a proper model of the polarizability along the imaginary axis which can be assembled from readily available quantities.

To represent polarizability at imaginary frequencies, we employ Pad\'e approximants which are rational functions denoted $[N/M]$, where $N$ and $M$ are degrees of polynomials in the numerator and denominator, respectively. \cite{Baker96} Because polarizabilities $\alpha_l(i\omega)$ depend only on the square of the frequency, $\omega^2$, we utilize Pad\'e approximants containing exclusively even powers in both numerator and denominator, given by
\begin{align}
\label{pade-gen}
    \alpha_l(i\omega) = \frac
    {\sum_{k=0}^N A_l^{(2k)}\,\omega^{2k}}
    {\sum_{k=0}^M B_l^{(2k)}\,\omega^{2k}}.
\end{align}
Without loss of generality, we can set $B_l^{(0)}=1$. To determine the remaining coefficients $A_l^{(2k)}$ and $B_l^{(2k)}$ we enforce the correct behavior of Pad\'e approximants in the $\omega\rightarrow0$ and $\omega\rightarrow\infty$ limits simultaneously. 

Consider first the small $\omega$ regime. When $|\omega|$ is not large, the polarizability can be expanded into the so-called Cauchy series which reads
\begin{align}
\label{cauchy-exp2}
    \alpha_l(\omega) = \alpha_l^{(0)} + \alpha_l^{(2)}\,\omega^2 + \alpha_l^{(4)}\,\omega^4 + \ldots
\end{align}
This expansion is valid for arbitrary frequency, i.e., both real and imaginary. Therefore, the coefficients $\alpha_l^{(2k)}$ can be obtained by calculating the corresponding polarizability for a set of real frequencies and fitting the results with the functional form~(\ref{cauchy-exp2}). Details of this procedure are given in Sec.~\ref{subsec:fitting}. The resulting coefficients $\alpha_l^{(2k)}$ can then be used without any change also along the imaginary axis, by simply performing the substitution $\omega\rightarrow i\omega$ in Eq.~(\ref{cauchy-exp2}). 

However, enforcing the correct behavior for small $\omega$ is insufficient to generate an accurate representation of the polarizability along the whole imaginary axis. To achieve this goal, we additionally need to anchor the Pad\'e approximants in the $\omega\rightarrow\infty$ limit. To this end, we derive the asymptotic expansion of $\alpha_l(i\omega)$ for large $\omega$. We start with the explicit expression for this quantity
\begin{align}
    \alpha_l(i\omega) = 2\sum_{n\neq0} \frac{\omega_n}{\omega_n^2+\omega^2}\,
    |\langle n| Q_{l0} | 0\rangle|^2,
\end{align}
and expand the component involving the excitation energies into an asymptotic series
\begin{align}
    \frac{\omega_n}{\omega_n^2+\omega^2} = \sum_{k=1}^\infty (-1)^{k-1}\,\omega_n^{2k-1}/\omega^{2k},
\end{align}
which is accurate for sufficiently large $\omega$. This leads to the formula
\begin{align}
    \alpha_l(i\omega) = \sum_{k=1}^\infty (-1)^{k-1}\,\frac{\zeta_l^{(2k)}}{\omega^{2k}},
\end{align}
where the coefficients are given by the general expression
\begin{align}
\label{def-zeta1}
    \zeta_l^{(2k)} = 2\sum_{n=0} \omega_n^{2k-1} |\langle n| Q_{l0} | 0\rangle|^2.
\end{align}
Note that in the latter formula we extended the sum over $n$ to include also the electronic ground state ($n=0$). This is allowed for each $k\geq1$ because $\omega_0=0$. To eliminate the cumbersome sum over states in the definition of $\zeta_l^{(2k)}$ we take advantage of the fact that the wavefunctions $|n\rangle$ with $n\geq0$ form a complete set. This enables us to eliminate the resolution of the identity from Eq.~(\ref{def-zeta1}) and rewrite the expansion coefficients in an equivalent form
\begin{align}
\label{def-zeta2}
    \zeta_l^{(2k)} = 2\langle 0| Q_{l0} \big( H - E_0 \big)^{2k-1}Q_{l0} | 0\rangle.
\end{align}

Let us consider the first asymptotic coefficient, namely $\zeta_l^{(2)}$. It can be simplified by using the formula $\big( H - E_0 \big)|0\rangle = 0$ and the fact that the potential in the Hamiltonian commutes with the multipole moment operators. This leads to 
\begin{align}
    \zeta_l^{(2)} = \langle 0| \big[ Q_{l0} , [ T, Q_{l0} ] \big] | 0\rangle,
\end{align}
where the electronic kinetic energy operator is $T=-\half\sum_i\nabla_i^2$. The nested commutator in the above formula can be evaluated explicitly
\begin{align}
\label{eq:defzeta1}
\begin{split}
    \zeta_l^{(2)}
    &=\langle 0|\sum_i\left(\nabla_i Q_{l0}\right)\cdot\left(\nabla_i Q_{l0}\right)|0\rangle \\
    &=\frac{4\pi}{2l+1}\sum_{m=-1}^1(-1)^m \\
    &\times\langle 0|\sum_i
    \left(\nabla_{i,m}\,r_i^l\,Y_{l0}(\hat{r}_i)\right)
    \left(\nabla_{i,-m}\,r_i^l\,Y_{l0}(\hat{r}_i)\right)
    |0\rangle.
\end{split}
\end{align}
The spherical components of the gradient operator $\nabla_i$ act on the function $r_i^l\, Y_{l0}(\hat{r}_i)$ according to \cite{Varshalovich}
\begin{align}
    \nabla_{i,0}\,r_i^l\,Y_{l0}(\hat{r}_i)&=l\sqrt{\frac{2l+1}{2l-1}}\,r_i^{l-1}\,Y_{l-1,0}(\hat{r}_i),\\
    \nabla_{i,\pm1}\,r_i^l\,Y_{l0}(\hat{r}_i)&=-\sqrt{\frac{l(l-1)}{2}}\sqrt{\frac{2l+1}{2l-1}}\,r_i^{l-1}\,Y_{l-1,\pm1}(\hat{r}_i).
\end{align}
Substituting these expressions into Eq.~(\ref{eq:defzeta1}) and using the relation $Y_{l-1,-m}=(-1)^mY_{l-1,m}^*$ yields
\begin{equation}
\label{eq:defzeta2}
\begin{split}
  \zeta_l^{(2)}
  &=\frac{4\pi\,l^2}{2l-1}\langle 0|\sum_i r_i^{2(l-1)}\,|Y_{l-1,0}(\hat{r}_i)|^2|0\rangle \\
  &+\frac{4\pi\,l(l-1)}{2l-1}\langle 0|\sum_i r_i^{2(l-1)}\,|Y_{l-1,1}(\hat{r}_i)|^2|0\rangle.
\end{split}
\end{equation}
Since the ground state of the nitrogen atom has $S$-type symmetry, only the spherically symmetric components of the operators in Eq.~(\ref{eq:defzeta2}) give nonvanishing contributions to the final result. The radial factor, $r_i^{2(l-1)}$, is already spherically symmetric, while the symmetric component of $|Y_{l-1,m}(\hat{r}_i)|^2$ is equal to its angular average
\begin{equation}
\label{eq:defzeta3}
\frac{1}{4\pi}\int|Y_{l-1,m}(\hat{r}_i)|^2\,\mathrm{d}\hat{r}_i=\frac{1}{4\pi},
\end{equation}
which follows from the normalization of spherical harmonics to unity. Replacing the angular terms in Eq.~(\ref{eq:defzeta2}) with this average, we arrive at the formula
\begin{equation}
  \label{eq:defzeta}
  \zeta_l^{(2)}=l\langle 0|\sum_i r_i^{2(l-1)}|0\rangle,
\end{equation}
where $r_i^{2(l-1)}=(x_i^2+y_i^2+z_i^2)^{l-1}$. In particular, in the special case $l=1$ (dipole polarizability) we obtain a simple formula, $\zeta_1^{(2)} = N$, where $N$ denotes the number of electrons in the system ($N=7$ in our particular case). This result is equivalent to the so-called Thomas--Reiche--Kuhn sum rule. \cite{Fano:68} In general, Eq.~(\ref{eq:defzeta}) shows that the leading coefficient in the asymptotic expansion of the $2^l$-pole polarizability for large (purely imaginary) frequencies is expressed through simple properties of the ground state. These quantities can be readily evaluated by using any wavefunction model employed in this work. We also derived the formula for the next asymptotic coefficient, $\zeta_l^{(4)}$, which is presented in the Supplementary Material. \cite{supp} Unfortunately, this expression proved to be significantly more complicated and difficult to evaluate accurately for many-electron systems. Furthermore, the complexity of the coefficients $\zeta_l^{(2k)}$ is expected to escalate further for $k>2$. Consequently, all higher asymptotic coefficients are omitted from the numerical calculations in this work.

Throughout this work, we shall use several first Cauchy coefficients, $\alpha_l^{(0)}$, $\alpha_l^{(2)}$, $\ldots$, $\alpha_l^{(4K-2)}$, and the first asymptotic coefficient $\zeta_l^{(2)}$ in the determination of the polarizability for imaginary frequencies. The Pad\'e approximant from Eq.~(\ref{pade-gen}) compatible with these data reads
\begin{align}
\label{pade-spec}
  \tilde{\alpha}_l^K(i\omega)=\frac
        {\sum_{k=0}^{K-1}A_l^{(2k)}\omega^{2k}+\zeta_l^{(2)}\omega^{2K}}
        {1+\sum_{k=1}^K B_l^{(2k)}\omega^{2k}+\omega^{2K+2}}.
\end{align}
Note that this form automatically exhibits the correct behavior for $\omega\rightarrow\infty$, namely $\tilde{\alpha}_l^K(i\omega)\rightarrow \zeta_l^{(2)}/\omega^2$. The remaining unknown coefficients $A_l^{(2k)}$ and $B_l^{(2k)}$ are found by expanding the above formula in a Taylor series around $\omega=0$ and comparing term-by-term with the Cauchy expansion up to $\omega^{4K-2}$. Once the Pad\'e approximants representing the polarizabilities are fully determined, the dispersion coefficients $C_6$, $C_8$, and $C_{10}$ are found by integration over $\omega$. The accuracy of this approach is validated in Sec.~\ref{subsec:pade} based on a set of benchmark calculations. Special emphasis is placed on the convergence of the results with respect to $K$.

\section{Numerical results}
\label{sec:numeric}

\subsection{Basis set considerations}
\label{subsec:basis}

In all calculations reported in this work, Gaussian basis sets are used to represent the atomic orbitals. The correlation-consistent Dunning family of basis sets (cc-pV$X$Z) from double- ($X=2$) to octuple-zeta ($X=8$) quality is available in the literature for nitrogen. \cite{dunning89,thorpe21} However, it is well-known that polarizability is highly sensitive to ``tails'' of the electronic density far from the nucleus and the standard cc-pV$X$Z basis sets are not designed to describe such properties. To improve the description of the long-range tail, they are usually augmented with additional sets of Gaussian functions with small exponents (diffuse functions). \cite{kendall92,woon94,thorpe21} Depending on the number of diffuse functions added, these basis sets are called singly-augmented (aug-cc-pV$X$Z), doubly-augmented (daug-cc-pV$X$Z), and so forth. For brevity, these conventional basis sets will be denoted by the symbols $X$Z, a$X$Z, da$X$Z, and so on, further in the text.

From the point of view of our applications, we found that these standard basis sets are suboptimal (in terms of accuracy to cost ratio) in calculation of the polarizabilities. For this reason, we modified the standard cc-pV$X$Z basis set family by adding a different set of diffuse functions. The remaining functions are unchanged in comparison with the parent basis sets. The number and type of Gaussian functions added to each basis set is given in Table~\ref{tab:basis}. We focus on singly- and doubly-augmented variants of the basis set. The latter are satisfactory in terms of accuracy and further augmentation leads to only minor improvements in the accuracy of the calculated polarizabilities. For brevity, we denote the new basis sets by the symbols a$X$Z(pol) and da$X$Z(pol), respectively.

\begin{table}
\caption{\label{tab:basis}
Number of diffuse functions included in the a$X$Z(pol) and da$X$Z(pol) basis sets. For example, the symbol 2$s$2$p$2$d$ means that two functions of $s$ symmetry ($l=0$), two functions of $p$ symmetry ($l=1$), and two functions of $d$ symmetry ($l=2$) were added.}
\begin{ruledtabular}
\begin{tabular}{cll}
 $X$ & 
 \multicolumn{1}{c}{a$X$Z(pol)} & 
 \multicolumn{1}{c}{da$X$Z(pol)} \\
 \hline\\[-1.2em]
 2 & 1$s$1$p$1$d$         & 2$s$2$p$2$d$ \\
 3 & 1$s$1$p$1$d$1$f$     & 2$s$2$p$2$d$2$f$ \\
 4 & 1$s$1$p$1$d$1$f$1$g$ & 2$s$2$p$2$d$2$f$2$g$ \\
 5 & 1$s$1$p$1$d$1$f$1$g$ & 2$s$2$p$2$d$2$f$2$g$ \\
 6 & 1$s$1$p$1$d$1$f$1$g$ & 2$s$2$p$2$d$2$f$2$g$ \\
\end{tabular}
\end{ruledtabular}
\end{table}

There are two major differences between the a$X$Z(pol)/da$X$Z(pol) and the standard a$X$Z/da$X$Z basis sets in terms of composition and optimization principle. First, as is evident from Table~\ref{tab:basis}, one or two sets of diffuse functions are added with angular momentum from $s$ to $g$ ($l=0$ to $l=4$). However, no functions with angular momentum higher than $l=4$ are ever included in the a$X$Z(pol)/da$X$Z(pol) basis sets. This is based on the numerical observation that such functions have a very small influence on the calculated polarizabilities (from dipole to octupole). This finding can be justified by considering a basic theory that can be used to calculate the polarizability, namely the coupled-perturbed Hartree--Fock method. In this method, the ground state wavefunction is represented by a single Slater determinant with occupied 1$s$, 2$s$, and 2$p$ orbitals. In calculations of polarizabilities, these orbitals are coupled to orbitals  with higher $l$ through the multipole moment operators present in the sum over states in Eq.~(\ref{alphaw1}). Using angular momentum algebra, one can show that for atoms the occupied $s$ and $p$ orbitals couple only up to $d$ orbitals in the case of dipole polarizability, up to $f$ orbitals for quadrupole polarizability, and up to $g$ orbitals for octupole polarizability. Higher angular momentum orbitals are unnecessary in this method. Of course, once a fully correlated method is used in the calculation, these simple arguments are no longer strictly valid, but since a single Slater determinant is a reasonably good approximation of the ground-state wavefunction of the nitrogen atom, one can expect the importance of diffuse orbitals with angular momentum higher than $l=4$ to be marginal, as observed numerically.

The second difference between the standard and a$X$Z(pol)/da$X$Z(pol) basis sets is the fact that the diffuse functions in the latter were directly optimized using static polarizability ($\omega=0$) as the target using full configuration interaction (FCI) method \cite{CI_method_1999} with frozen 1$s$ orbital (five active electrons). However, in contrast to the atomic energy, the polarizability is not a variational quantity and hence cannot be optimized directly with respect to basis set parameters without access to highly-accurate reference data (which is not available in our case). To avoid this problem and to define a suitable optimization target we introduce a Hylleraas-like functional which is bound from below. This functional reads
\begin{align}
    \mathcal{H}_l = 4\langle r| Q_{l0} |0\rangle - 2\langle r| H - E_0 |r\rangle,
\end{align}
where $|r\rangle$ denotes the first-order response function that is formally defined as
\begin{align}
    |r\rangle = \big( H - E_0 \big)^{-1} Q_{l0} |0\rangle,
\end{align}
with the orthogonality constraint $\langle0|r\rangle = 0$. As this functional is quadratic in the response function, one can show by direct calculation that it is bounded from below with respect to variations in $|r\rangle$. Moreover, the minimum of the functional $\mathcal{H}_l$ corresponds to the $2^l$-pole polarizability as defined in Eqs.~(\ref{alphaw1})~and~(\ref{alphaw2}). In our calculations, both $|0\rangle$ and $|r\rangle$ are represented by the FCI expansions, and the functionals $\mathcal{H}_l$ are minimized with respect to the exponents of the diffuse Gaussian functions added to the basis. More precisely, the exponents of the $s$, $p$, and $d$ functions are optimized using the $\mathcal{H}_1$ functional. These exponents are then fixed and the exponents of the additional $f$ and $g$ functions are optimized using the $\mathcal{H}_2$ and $\mathcal{H}_3$ functionals, respectively. The detailed composition of the optimized basis sets, including values of the exponents of the additional Gaussian functions, is provided in the Supplementary Material. \cite{supp} Note that the basis set design principle adopted in this work is similar to the ``calendar'' basis sets~\footnote{the term ``calendar'' is a pun based on naming the aug- basis sets as ``August'' and using earlier months to denote smaller basis sets with less diffuse functions} (jul-cc-pV$X$Z, jun-cc-pV$X$Z, etc.), where the least important diffuse functions for a given quantity are systematically eliminated (see Ref.~\onlinecite{papajak11} and references therein).

\begin{figure}
    \hspace{-0.75cm}\includegraphics[width=0.85\linewidth]{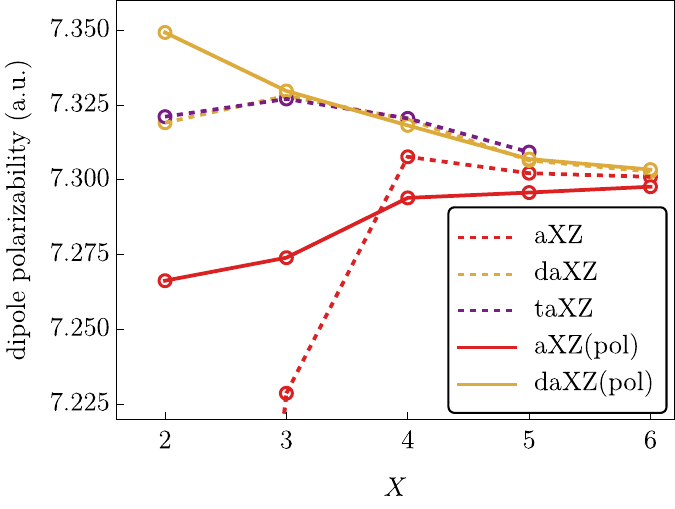}\vspace{0.25cm} \\
    \hspace{-0.75cm}\includegraphics[width=0.85\linewidth]{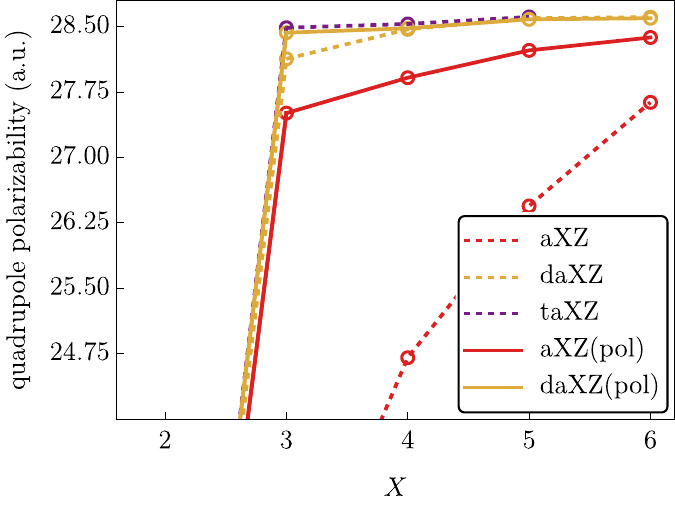}\vspace{0.25cm} \\
    \hspace{-0.75cm}\includegraphics[width=0.85\linewidth]{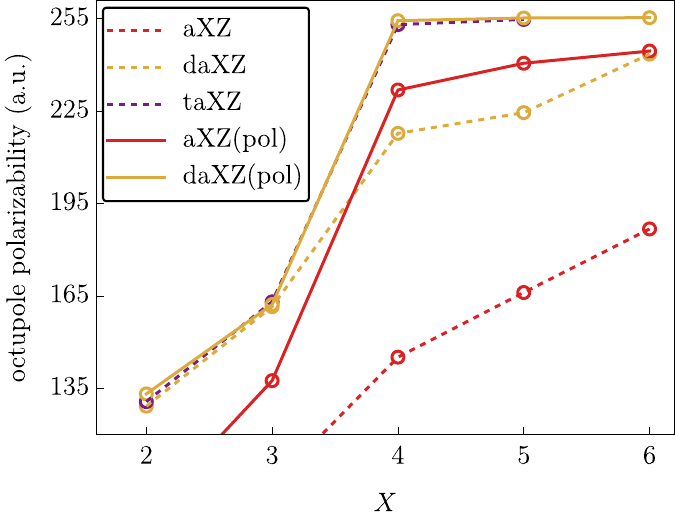} \\
    \caption{Static dipole (upper panel), quadrupole (middle panel) and octuple (lower panel) polarizabilities calculated using the frozen-core ($1\mathrm{s}^2$) CCSDT method using the a$X$Z, da$X$Z, ta$X$Z, a$X$Z(pol), and da$X$Z(pol) basis set families as a function of $X=2,\ldots,6$. The results for the ta6Z basis set are not available due to prohibitive computational costs.}
    \label{fig:basis-set-tests}
\end{figure}

To illustrate the performance of the new basis sets, we compare the results of the static polarizabilities calculated with the a$X$Z(pol)/da$X$Z(pol) basis sets against those obtained with the standard Dunning basis sets up to the triple-augmentation level, see Fig.~\ref{fig:basis-set-tests}. To prove that the new basis sets are not biased towards the level of theory they were optimized at (FCI), the data in Fig.~\ref{fig:basis-set-tests} are based on the frozen-core ($1s^2$) coupled cluster method with single, double and triple excitations (CCSDT). \cite{noga87,scuseria88} The CCSDT calculations reported in this work were conducted using a locally modified version of the {\sc NWChem} program package. \cite{nwchem} The publicly-available version of this program supports the calculation of only dipole polarizabilities and hence it was modified by inserting the matrix elements of higher-order electric moments to enable calculation of quadrupole and octupole polarizabilities in this work.

Starting with the dipole polarizability from Fig.~\ref{fig:basis-set-tests}, we see that there is no significant difference between the da$X$Z(pol) and da$X$Z basis sets for $X>2$. Moreover, triple augmentation changes the results to a very small degree as the curves for da$X$Z and ta$X$Z in Fig.~\ref{fig:basis-set-tests} are nearly on top of each other. This proves that the double augmentation level is entirely sufficient for dipole polarizabilities, independently of the basis set family. Singly-augmented basis sets a$X$Z(pol) and a$X$Z offer a similar accuracy level, but not satisfactory from the point of view of the present work. The only significant difference between a$X$Z(pol) and a$X$Z in the case of dipole polarizability is somewhat more regular convergence in the case of the former. Similar conclusions as for the dipole polarizability hold also for the quadrupole polarizability, with the exception that da$X$Z(pol) converges to the basis set limit somewhat more regularly than da$X$Z. The major deviations between the new and standard basis set families are observed for octupole polarizability. First, da$X$Z and ta$X$Z yield substantially different results, suggesting that the augmentation functions used in da$X$Z are not optimal. Next, results obtained within da$X$Z behave irregularly; for example, the change in the values from da4Z to da5Z is \emph{smaller} than that from da5Z to da6Z, contrary to what one would expect from a systematic sequence of basis sets. No such artifacts are observed for the a$X$Z(pol) and da$X$Z(pol) basis sets. Moreover, the results obtained within the da$X$Z(pol) basis agree very well with the standard ta$X$Z basis. This increase of accuracy and stability of the results is a significant advantage; for example, the FCI and CCSDT calculations for nitrogen are feasible within the da$X$Z(pol) basis sets, but prohibitively expensive using ta$X$Z due to the increased size of the latter.

\subsection{Accuracy of the Cauchy coefficients obtained by fitting}
\label{subsec:fitting}

In the calculations reported in this work, there are several sources of error and each of them must be strictly controlled. The most basic source of error is the incompleteness of both the basis set and the adopted wavefunction model. These errors are addressed and minimized in Sec.~\ref{subsec:cauchy}. However, another source of error, which is potentially significant, is the fact that the Cauchy coefficients in the expansion of the polarizability cannot be obtained directly using most electronic structure methods. While for closed-shell systems direct determination of the Cauchy coefficients has been implemented for some variants of the coupled cluster theory, \cite{Hattig:1997} we are not aware of any such implementation applicable to open-shell atoms. As a result, direct determination of the Cauchy coefficients for nitrogen is possible at present only at the FCI level of theory using in-house program {\sc Hector} written by one of us (M.P.). \cite{hector} In Appendix~\ref{sec:appb} we describe the computational procedure implemented for direct calculation of the Cauchy coefficients. For any other method used in this work, we employ a fitting procedure in which the polarizability is calculated for a set of real frequencies. Next, a polynomial expansion is fitted to the set of available points using the least-squares method which enables us to find the Cauchy coefficients.

To test the reliability of this procedure, we used dipole polarizability of the nitrogen atom obtained using the frozen-core FCI method within the da4Z(pol) basis set as an example. The Cauchy coefficients were determined in two ways: (i) directly via the FCI calculations as described in Appendix~\ref{sec:appb} and (ii) through the fitting procedure based on a set of $28$ frequencies lying within the interval $\omega\in[0.0,0.2]$ (in atomic units). The full list of frequencies employed in this procedure is given in the Supplementary Material. \cite{supp} Further extension of this set by adding some points in-between the listed frequencies leads to only minor changes in the final results. 

As both approaches to the determination of the Cauchy coefficients use the same basis set and wavefunction model, they should, in principle, give the same results. However, this would be true only if one assumes infinite precision of the calculated results. In practice, each calculated point carries a numerical error which can accumulate in the fitting procedure. We conservatively assume that the results can be converged to $10$ significant digits and truncate the raw results at this accuracy level before the fitting procedure is carried out. In Table~\ref{tab:fitting} we compare the Cauchy coefficients, defined by Eq.~(\ref{cauchy-exp2}) with $l=1$, obtained using the direct method with those from the fitting procedure including even powers of $\omega$ up to $\omega^{30}$. Any deviation between the two sets of results is attributed solely to the inherent error of the fitting procedure. As shown in Table~\ref{tab:fitting}, the first two Cauchy coefficients ($\alpha_1^{(0)}$ and $\alpha_1^{(2)}$) obtained by fitting are essentially identical to the reference values. A slightly larger deviation between the direct method and fitting is observed for the next coefficient ($\alpha_1^{(4)}$), but this discrepancy is below $10^{-4}$ in relative terms and hence not a cause for concern from the point of view of this work. Unfortunately, the higher-order Cauchy coefficients are progressively less accurate. The relative error of the $\alpha_1^{(6)}$ coefficient is about 2\% which may be acceptable in some contexts, but must be treated with care. The last coefficient shown in Table~\ref{tab:fitting}, namely $\alpha_1^{(8)}$, is clearly unphysical due to a wrong sign (all Cauchy coefficients are positive by definition). In summary, from the point of view of the present work the fitting procedure is trustworthy only for the first three Cauchy coefficients. Higher-order coefficients must be evaluated by the direct method.

\begin{table}
\caption{\label{tab:fitting}
Comparison of the Cauchy coefficients $\alpha_1^{(2n)}$ for the nitrogen atom obtained using the direct method, see Appendix~\ref{sec:appb}, and by fitting dynamic polarizabilities calculated for a set of 28 real frequencies. Both sets of results were obtained using the frozen-core ($1s^2$) FCI method within the da4Z(pol) basis set.
}
\begin{ruledtabular}
\begin{tabular}{ccc}
 coefficient & direct method & fitting \\
 \hline\\[-1.2em]
 $\alpha_1^{(0)}$ & 7.31503 & \phantom{$-$}7.31503 \\
 $\alpha_1^{(2)}$ & 27.1103 & \phantom{$-$}27.1103 \\
 $\alpha_1^{(4)}$ & 142.492 & \phantom{$-$}142.479 \\
 $\alpha_1^{(6)}$ & 863.503 & \phantom{$-$}883.717 \\
 $\alpha_1^{(8)}$ & 5573.39 & $-$9523.98 \\
\end{tabular}
\end{ruledtabular}
\end{table}

\subsection{Accuracy of the Pad\'e approximants}
\label{subsec:pade}

\begin{table*}
\caption{\label{tab:pade}
Dispersion coefficients $\tilde{C}_n^K$ for the helium dimer calculated using $2^l$-pole dynamic polarizabilities $\tilde{\alpha}_l^K(i\omega)$ represented as Pad\'e approximants of degree $K$, see Eq.~(\ref{pade-spec}). The corresponding results obtained using the sum-over-states (SOS) approach are given in the last row ($C_n^\mathrm{SOS}$). The symbols $\Delta_n$ denote relative errors with respect to the SOS results; for example, $\Delta_6=(\tilde{C}_6^K-C_6^\mathrm{SOS})/C_6^\mathrm{SOS}$.}
\begin{ruledtabular}
\begin{tabular}{cd{1.6}d{1.6}d{2.5}d{1.6}d{3.4}d{1.6}}
$K$ &
\multicolumn{1}{c}{$\tilde{C}_6^{K}$} & \multicolumn{1}{c}{$\Delta_6$} &
\multicolumn{1}{c}{$\tilde{C}_8^{K}$} & \multicolumn{1}{c}{$\Delta_8$} &
\multicolumn{1}{c}{$\tilde{C}_{10}^{K}$} & \multicolumn{1}{c}{$\Delta_{10}$} \\
\hline
1   & 1.553431 & 6.3\cdot10^{-2} & 15.06751 & 6.7\cdot10^{-2} & 195.8369 & 6.7\cdot10^{-2} \\
2   & 1.478197 & 1.2\cdot10^{-2} & 14.27215 & 1.1\cdot10^{-2} & 185.2610 & 8.9\cdot10^{-3} \\
3   & 1.464670 & 2.4\cdot10^{-3} & 14.14448 & 1.6\cdot10^{-3} & 183.8311 & 1.1\cdot10^{-3} \\
4   & 1.461392 & 1.9\cdot10^{-4} & 14.12390 & 1.2\cdot10^{-4} & 183.6378 & 9.0\cdot10^{-5} \\
5   & 1.461151 & 2.3\cdot10^{-5} & 14.12240 & 1.6\cdot10^{-5} & 183.6236 & 1.2\cdot10^{-5} \\
6   & 1.461125 & 4.6\cdot10^{-6} & 14.12223 & 3.5\cdot10^{-6} & 183.6218 & 2.8\cdot10^{-6} \\
\hline
SOS & 1.461118 &                & 14.12218 &                & 183.6213 &                \\
\end{tabular}
\end{ruledtabular}
\end{table*}

\begin{table*}
\caption{\label{tab:pade-mod}
Same as Table~\ref{tab:pade}, except that the $2^l$-pole dynamic polarizabilities are represented as Pad\'e approximants generated using modified sets of Cauchy coefficients (see text). The exact values of the dispersion coefficients \cite{Yan1996} are given in the last row ($C_n^\mathrm{exact}$). The symbols $\Delta_n$ denote relative errors with respect to the exact results; for example, $\Delta_6=(\tilde{C}_6^K-C_6^\mathrm{exact})/C_6^\mathrm{exact}$.
}
\begin{ruledtabular}
\begin{tabular}{cd{1.6}d{1.6}d{2.5}d{1.6}d{3.4}d{1.6}}
$K$ &
\multicolumn{1}{c}{$\tilde{C}_6^{K}$} & \multicolumn{1}{c}{$\Delta_6$} &
\multicolumn{1}{c}{$\tilde{C}_8^{K}$} & \multicolumn{1}{c}{$\Delta_8$} &
\multicolumn{1}{c}{$\tilde{C}_{10}^{K}$} & \multicolumn{1}{c}{$\Delta_{10}$} \\
\hline
1     & 1.553118 & 6.3\cdot10^{-2} & 15.06369 & 6.7\cdot10^{-2} & 195.9895 & 6.7\cdot10^{-2} \\
2     & 1.477904 & 1.2\cdot10^{-2} & 14.26964 & 1.1\cdot10^{-2} & 185.4735 & 9.7\cdot10^{-3} \\
3     & 1.464381 & 2.3\cdot10^{-3} & 14.14233 & 1.7\cdot10^{-3} & 184.0631 & 2.0\cdot10^{-3} \\
4     & 1.461104 & 8.6\cdot10^{-5} & 14.12183 & 2.8\cdot10^{-4} & 183.8710 & 9.8\cdot10^{-4} \\
5     & 1.460863 &-7.8\cdot10^{-5} & 14.12034 & 1.8\cdot10^{-4} & 183.8570 & 9.0\cdot10^{-4} \\
6     & 1.460837 &-9.6\cdot10^{-5} & 14.12017 & 1.6\cdot10^{-4} & 183.8554 & 8.9\cdot10^{-4} \\[0.5ex]
mean($K\ge4$)& 1.460935 &-2.9\cdot10^{-5} & 14.12078 & 2.1\cdot10^{-4} & 183.8611 & 9.3\cdot10^{-4} \\
\hline
exact & 1.460978 &                 & 14.11786 &                 & 183.6911 &                 \\
\end{tabular}
\end{ruledtabular}
\end{table*}

Another source of error in our calculation is related to the representation of the polarizability for imaginary frequencies using the Pad\'e approximants. Clearly, this error affects only the dispersion coefficients reported in this work. It is obvious that the rational functions~(\ref{pade-gen}) are able to approximate any continuous function on the interval $\omega\in[0,\infty)$. However, it is necessary to test how the achieved accuracy depends on the number of terms and whether it is sufficient for our purposes. To answer this question, we carried out a systematic benchmark study for the helium atom. The use of helium as a testbed is motivated by the following observations:
\begin{enumerate}
    \item polarizabilities at imaginary frequencies have qualitatively almost the same shape for all light atoms;
    \item for helium the dispersion coefficients $C_n$ can be evaluated directly using the sum-over-states (SOS) approach, see Ref.~\onlinecite{Przybytek:2008}, which enables us to cross-check the approach based on Pad\'e approximants; this is not possible (or at least prohibitively costly) for many-electron atoms such as nitrogen;
    \item accurate reference data for the dispersion coefficients $C_n$ of helium exist in the literature, \cite{Yan1996} enabling an independent verification.
\end{enumerate}

In the calculations based on Casimir--Polder formulas, Eqs.~(\ref{c6cp})--(\ref{c10cp}), we use the polarizabilities $\tilde{\alpha}_l^K(i\omega)$ represented as Pad\'e approximants of degree $K$, see Eq.~(\ref{pade-spec}). The dispersion coefficients obtained using $\tilde{\alpha}_l^K(i\omega)$ for a given $K$ are denoted by the symbols $\tilde{C}_6^{K}$, $\tilde{C}_8^{K}$, and $\tilde{C}_{10}^{K}$. The results for $K=1,\ldots,6$ are presented in Table~\ref{tab:pade} together with the corresponding data obtained using the SOS method. All results given in Table~\ref{tab:pade} were obtained at the FCI level of theory within the doubly-augmented correlation-consistent basis set of the 6-zeta quality taken from Ref.~\onlinecite{Cencek:2012}.

As the calculations based on the SOS approach and on the Pad\'e approximants employ the same electronic structure method and the same basis set, they are identical in the limit $K\rightarrow\infty$. Therefore, the difference between the two sets of results for a given $K$ is attributed solely to the incompleteness of the Pad\'e expansion. From Table~\ref{tab:pade} we see that the convergence with respect to $K$ is rapid and it is possible to achieve levels of accuracy as high as a few parts per million (with $K=6$) in relative terms. From the point of view of the present work, choosing at least $K=4$ is required to reduce the relative error below 0.1\% with respect to the SOS result.

When using a basis set of 6-zeta quality, the dispersion coefficients for helium calculated via the SOS method show relative deviations from the exact values from Ref.~\onlinecite{Yan1996} of $9.6\cdot10^{-5}$, $3.1\cdot10^{-4}$, and $-3.8\cdot10^{-4}$ for $C_6$, $C_8$, and $C_{10}$, respectively. Within the SOS approach, these errors can be reduced only by employing larger basis sets. In contrast, the Casimir--Polder approach allows for a different strategy: we can use polarizabilities represented as Pad\'e approximants generated from a set of Cauchy coefficients, where selected values of $\alpha_l^{(2k)}$ computed within a given finite basis set are replaced by more accurate ones determined through other approaches. For helium, as an example, the \(\alpha _{l}^{(0)}\) coefficients can be substituted with the exact static polarizabilities \(\alpha_l(0)\) taken from the literature. \cite{Yan1996} We found, however, that replacing $\alpha_l^{(0)}$ while leaving the remaining Cauchy coefficients intact may lead to Pad\'e approximants $\tilde{\alpha}_l^K(i\omega)$ with singularities at positive values of $\omega$ when $K\ge3$. To circumvent this unphysical behavior, we scale the higher coefficients, $\alpha_l^{(2k)}$ with $k>0$, by the ratio $\alpha_l^{(0),\text{exact}}/\alpha_l^{(0),\text{approx}}$, where $\alpha_l^{(0),\text{exact}}$ is the exact static polarizability and $\alpha_l^{(0),\text{approx}}$ is the value obtained with a finite basis set. By applying this scaling, only the octupole polarizabilities $\tilde{\alpha}_3^K(i\omega)$ still exhibit singularities for $K\ge5$.

The dispersion coefficients $\tilde{C}_6^K$, $\tilde{C}_8^K$, and $\tilde{C}_{10}^K$, $K=1,\ldots,6$, obtained using $\tilde{\alpha}_l^K(i\omega)$ generated from modified sets of Cauchy coefficients, are presented in Table~\ref{tab:pade-mod}. Since the Pad\'e approximants $\tilde{\alpha}_3^K(i\omega)$ are singular for $K\ge5$, the $\tilde{\alpha}_3^4(i\omega)$ variant was employed to calculate $\tilde{C}_{10}^5$ and $\tilde{C}_{10}^6$. We see that for small $K$ the convergence to the exact values of $C_n$ is rapid, whereas for $K\ge4$ it slows down significantly while remaining monotonic. Although observed for helium, such monotonic convergence may not be the case for other atoms. Therefore, as the recommended values of dispersion coefficients from the Casimir--Polder approach, $\tilde{C}_n$, we take the mean of the results obtained in the range of $K$ where stabilization is observed [reported as the 'mean($K\ge4$)' entry in Table~\ref{tab:pade-mod}]. In the case of $\tilde{C}_6$ and $\tilde{C}_8$, the exact values are reproduced with a relative error of about $10^{-4}$. More importantly, the inclusion of accurate $\alpha_l^{(0)}$ coefficients in the construction of Pad\'e approximants shifts the results closer to the exact values: the absolute relative error of the final $\tilde{C}_6$ value is reduced from $9.6\cdot10^{-5}$ to $2.9\cdot10^{-5}$, and for $\tilde{C}_8$ from $3.1\cdot10^{-4}$ to $2.1\cdot10^{-4}$. A different situation is observed for $\tilde{C}_{10}$. In this case, the use of modified Cauchy coefficients actually worsens the agreement with the exact value, increasing the absolute relative error from $3.8\cdot10^{-4}$ to $9.3\cdot10^{-4}$. This may be attributed to the inability to construct nonsingular Pad\'e approximants for $\tilde{\alpha}_3^5(i\omega)$ and $\tilde{\alpha}_3^6(i\omega)$, which results in suboptimal $\tilde{C}_{10}^5$ and $\tilde{C}_{10}^6$.

The procedure described above---centered on combining accurate lower Cauchy coefficients with higher coefficients calculated within a finite basis set---will be subsequently applied to the calculations for the nitrogen atom.

\subsection{Calculation of the Cauchy coefficients}
\label{subsec:cauchy}

From the point of view of applications mentioned in the introduction, the first four Cauchy coefficients are the most important in practice. For example, for frequencies corresponding to wavelengths of helium-neon laser ($\approx 633\,$nm) and longer, inclusion of the first four Cauchy coefficients ($\alpha_1^{(0)}$, $\alpha_1^{(2)}$, $\alpha_1^{(4)}$, $\alpha_1^{(6)}$) is sufficient to determine the polarizability with relative accuracy of roughly $10\,$ppm. Moreover, the last coefficient ($\alpha_1^{(6)}$) is required with accuracy of only about $7-8\%$ which is not difficult to achieve. Therefore, we first focus on theoretical calculation of the $\alpha_1^{(0)}$, $\alpha_1^{(2)}$, $\alpha_1^{(4)}$ coefficients at the limits of accuracy that can be achieved at present.

To this end, we employ a composite approach in which we combine several methods and include a set of progressively smaller corrections accounting for various physical effects. Let us denote a quantity being calculated by the symbol $Y$. It is calculated as a sum of four contributions, namely
\begin{align}
\label{composite}
    Y = Y_{\rm{fc-FCI}} + \delta Y_{\rm ae-CCSDT} + \delta Y_{\rm ae-FCI} + \delta Y_{\rm rel1}.
\end{align}
The first term, $Y_{\rm{fc-FCI}}$, is the dominant valence contribution (with five active electrons) calculated using the FCI method. The second term, $\delta Y_{\rm ae-CCSDT}$, is a correction that accounts for the correlation effects due to the $1s^2$ core orbitals which are inactive in the $Y_{\rm{fc-FCI}}$ term. It is defined as the difference between all-electron and valence-only CCSDT calculations in the same basis set. The $\delta Y_{\rm ae-FCI}$ term is defined analogously but includes only post-CCSDT core correlation effects, i.e., it accounts for quadruple and higher core-core and mixed core-valence excitations in the CC theory. The reason for this division is the fact that $\delta Y_{\rm ae-CCSDT}$ can usually be calculated in a larger basis set than $\delta Y_{\rm ae-FCI}$, leading to decreased uncertainties. Finally, the $\delta Y_{\rm rel1}$ term collects relativistic corrections to the quantity $Y$. The details of the procedure used to calculate $\delta Y_{\rm rel1}$ are given further in the text.

In Table~\ref{tab:valencefci} we report results of the calculations of the dominant contribution to the Cauchy coefficients, namely $Y_{\rm{fc-FCI}}$, within the da$X$Z(pol) basis set family, $X=2,\ldots,6$. To minimize the basis set incompleteness error, we employ the extrapolation scheme based on the Riemann zeta function from Ref.~\onlinecite{lesiuk19Riemann}. To estimate the uncertainty of the extrapolated results, we used the procedure developed in Ref.~\onlinecite{lang25random} based on a series of random walks with ten million independent samples. All error bars given here refer to the $2\sigma$ confidence interval ($\approx 95\%$ confidence level). The same protocol is adopted for the basis set extrapolation and the corresponding uncertainty estimation throughout the present work unless explicitly stated otherwise. We also tested an alternative extrapolation method proposed by Helgaker and collaborators \cite{Halkier1998,Helgaker2008} but found only minor differences in the results that do not exceed the assigned error bars. The corresponding $Y_{\rm{fc-FCI}}$ results for the remaining coefficients, $\alpha_2^{(2k)}$ and $\alpha_3^{(2k)}$, are given in the Supplementary Material. \cite{supp}

\begin{table}
\caption{\label{tab:valencefci}
Cauchy coefficients $\alpha_1^{(2k)}$ for the nitrogen atom calculated using the frozen-core FCI method and the da$X$Z(pol) basis sets family. Results extrapolated to the complete basis set limit and their error estimates are given in the last row. Numbers in the parentheses denote the uncertainty estimates at the last digit(s).
}
\begin{ruledtabular}
\begin{tabular}{cd{8}d{7}d{7}}
$X$ & 
\multicolumn{1}{c}{$\alpha_1^{(0)}$} &
\multicolumn{1}{c}{$\alpha_1^{(2)}$} &
\multicolumn{1}{c}{$\alpha_1^{(4)}$} \\
\hline\\[-1.2em]
2 & 7.345341 & 27.38318 & 145.7249 \\
3 & 7.326389 & 27.14779 & 142.8741 \\
4 & 7.315028 & 27.11028 & 142.4919 \\
5 & 7.303870 & 27.06622 & 142.1385 \\
6 & 7.300354 & 27.03894 & 141.9554 \\
\hline\\[-1.2em]
$\infty$ & 7.2949(71) & 26.997(23) & 141.671(48) \\
\end{tabular}
\end{ruledtabular}
\end{table}

The next contribution, $\delta Y_{\rm ae-CCSDT}$, was calculated using the CCSDT method with the daC$X$Z(pol) basis set family, $X=2,\ldots,5$. These basis sets were obtained by supplementing the valence-only da$X$Z(pol) basis sets used in the $Y_{\rm{fc-FCI}}$ calculations with the sets of ``tight'' functions taken from the cc-pCV$X$Z family developed in Ref.~\onlinecite{woon95}. The reason for this modification is to increase the flexibility of the basis in regions close to the atomic nuclei, which is essential for an accurate description of the core correlation effects. The raw results of the $\delta Y_{\rm ae-CCSDT}$ correction to $\alpha_l^{(2k)}$ are given in the Supplementary Material, \cite{supp} while the final extrapolated results for $\alpha_1^{(2k)}$ and the corresponding uncertainty estimates are included in Table~\ref{tab:summarya1n}.

\begin{table}
\caption{\label{tab:summarya1n}
Summary of the results obtained for the Cauchy coefficients $\alpha_1^{(2k)}$ of the nitrogen atom. Numbers in the parentheses denote the uncertainty estimates at the last digit(s).}
\begin{ruledtabular}
\begin{tabular}{cd{10}d{9}d{8}}
contribution & 
\multicolumn{1}{c}{$\alpha_1^{(0)}$} & 
\multicolumn{1}{c}{$\alpha_1^{(2)}$} & 
\multicolumn{1}{c}{$\alpha_1^{(4)}$} \\
\hline\\[-1.2em]
$Y_\mathrm{fc-FCI}$          &       7.2949(71) &       26.997(23) &      141.671(48) \\
$\delta Y_\mathrm{ae-CCSDT}$ &      -0.0366(19) &       -0.190(13) &        -1.06(14) \\
$\delta Y_\mathrm{ae-FCI}$   &      0.00057(23) &       0.0069(26) &        0.045(12) \\
$\delta Y_\mathrm{rel1}$     &    -0.003630(17) &     -0.05707(24) &      -0.5862(76) \\
\hline\\[-1.2em]
total                        &       7.2552(74) &       26.756(27) &       140.07(15) \\
\end{tabular}
\end{ruledtabular}
\end{table}

The next correction, $\delta Y_{\rm ae-FCI}$, is particularly computationally expensive and we managed to evaluate it using only the daC$2$Z(pol) and daC$3$Z(pol) basis sets. Fortunately, this term is also tiny and hence it does not have to be calculated with precision comparable to the previous contributions. The final value of the $\delta Y_{\rm ae-FCI}$ correction to each Cauchy coefficient was obtained by Riemann extrapolation from the $X=2,3$ basis-set pair. However, as the results from only two basis sets are available, we cannot perform the uncertainty estimation procedure from Ref.~\onlinecite{lang25random} which requires data from at least three consecutive basis sets. Therefore, the uncertainty was estimated more conservatively by taking the difference between the extrapolated result and the value obtained within the $X=3$ basis set. The final results are given in Table~\ref{tab:summarya1n} in the case of $\alpha_1^{(2k)}$  and in the Supplementary Material for the quadrupole and octupole Cauchy coefficients. \cite{supp}

The final contribution, $\delta Y_{\rm rel1}$, accounts for relativistic corrections to Cauchy coefficients and requires some discussion. First, we consider only terms proportional to $1/c^2$, where $c$ is the speed of light. As shown further in the text, even the leading-order $1/c^2$ corrections are small (albeit non-negligible), and hence the inclusion of $1/c^3$ and higher-order ones is not necessary. Second, regarding the $1/c^2$ relativistic corrections arising from the Breit--Pauli Hamiltonian, \cite{BeSal,Pachucki2004} they can be divided into spin-free (scalar) and spin-dependent terms, and simultaneously into one- and two-electron corrections. It is known that the spin-dependent terms for atomic $S$ states either vanish or can be rewritten as a combination of spin-free two-electron ones. The two-electron scalar corrections for light atoms are at least several times smaller than one-electron contributions, as illustrated in calculations for helium, \cite{puchalski20} lithium, \cite{puchalski11} neon, \cite{lesiuk20} or even argon. \cite{lesiuk23} Consequently, the inclusion of two-electron corrections in this work is unnecessary, as their values would drown in comparison with the uncertainties of other non-relativistic contributions. Therefore, in the $\delta Y_{\rm rel1}$ term, we include only scalar one-electron relativistic effects. They can be calculated either perturbatively starting with the Breit--Pauli Hamiltonian or using the scalar second-order Douglas--Kroll--Hess (DKH2) method. \cite{Douglas:1974,Hess:1986,Jansen:1989} We found the latter method to be more appealing from a technical point of view and adopted it for the calculation of $\delta Y_{\rm rel1}$. Note that in comparison with the pure perturbative approach, the DKH2 method includes some extra relativistic terms of the order of $1/c^4$ and higher, but their magnitude is negligible. The $\delta Y_{\rm rel1}$ term was calculated using the all-electron CCSDT method with the da$X$Z(pol) basis sets, $X=3,\ldots,5$. However, to further increase the flexibility of basis sets, they were uncontracted in these calculations. The $\delta Y_{\rm rel1}$ correction is computed as the difference between DKH2+CCSDT and non-relativistic CCSDT results within the same basis set. The raw results of the $\delta Y_{\rm rel1}$ correction to $\alpha_l^{(2k)}$ are given in the Supplementary Material, \cite{supp} while in Table~\ref{tab:summarya1n} we include the final results for $\alpha_1^{(2k)}$ extrapolated to the complete basis set limit using the same protocol as described above for the $Y_{\rm fc-FCI}$ contribution.

In Table~\ref{tab:summarya1n}, we provide a summary of calculations for $\alpha_1^{(2k)}$ coefficients including all four contributions defined in Eq.~(\ref{composite}) along with the corresponding uncertainty estimates. An analogous summary for $\alpha_2^{(2k)}$ and $\alpha_3^{(2k)}$ coefficients is provided in the Supplementary Material. \cite{supp} The total uncertainty estimates are assembled from those of the individual components using the standard error propagation rules, i.e., by summing squares of uncertainties for each contribution and taking the square root of the sum. Clearly, the valence-only term $Y_{\rm{fc-FCI}}$ brings by far the most sizeable contribution to the calculated quantities and, in most cases, the uncertainty of this term dominates the overall error budget. We see that the $\delta Y_{\rm ae-FCI}$ term is particularly small and it could be neglected without a drastic increase in the overall error. This conclusion is fortunate because the $\delta Y_{\rm ae-FCI}$ term also carries the largest \emph{relative} uncertainty. The relativistic corrections $\delta Y_{\rm rel1}$ are also small, but it is not recommended to neglect them as this would increase the total uncertainty by a significant margin. All in all, the $\alpha_1^{(2k)}$ coefficients are determined with a relative uncertainty of about $0.1\%$ or better. Slightly larger uncertainties, up to roughly 0.5\% are found for the first two quadrupole and octupole Cauchy coefficients, see the Supplementary Material, \cite{supp} while for the third coefficient the uncertainties reach a few percent.

\subsection{Calculation of the dispersion coefficients}
\label{subsec:cdisp}

Knowing the Cauchy coefficients determined in the previous section, we are in a position to calculate the dispersion coefficients $C_6$, $C_8$, and $C_{10}$ for two nitrogen atoms using the Casimir--Polder formulas given in Eqs.~(\ref{c6cp})--(\ref{c10cp}), where the dynamic polarizabilities at imaginary frequencies are represented via Pad\'e approximants of the form given in Eq.~(\ref{pade-spec}). We follow the protocol discussed in Sec.~\ref{subsec:pade}: to generate the Pad\'e approximants, we utilize a baseline set of Cauchy coefficients calculated using the frozen-core FCI method within the da6Z(pol) basis set, replacing the leading coefficient and scaling the higher-order ones as detailed below. This baseline set includes $\alpha_1^{(2k)}$, $\alpha_2^{(2k)}$, and $\alpha_3^{(2k)}$ with $k$ ranging from 0 to 17. Their values are listed in the Supplementary Material. \cite{supp} Constructing the Pad\'e approximants according to Eq.~(\ref{pade-spec}) also requires the asymptotic coefficients $\zeta_l^{(2)}$. As discussed in Sec.~\ref{sec:theory}, the value $\zeta_1^{(2)}=7$ is exact, whereas $\zeta_2^{(2)}=24.367(38)$ and $\zeta_3^{(2)}=198.68(53)$ are calculated in the Supplementary Material. \cite{supp}

The size of the baseline set of Cauchy coefficients enables the generation of Pad\'e approximants $\tilde{\alpha}_l^K(i\omega)$ of degrees up to $K=9$. We found that replacing either the two or three lowest-order coefficients in the baseline set with accurate values obtained in Sec.~\ref{subsec:cauchy} leads to singular approximants for larger $K$, particularly in the case of the octupole polarizability. This behavior can be attributed to two factors. First, as $K$ increases, Pad\'e approximants become increasingly sensitive to the numerical precision of the Cauchy coefficients, whereas our accurate values are known to at most five significant digits. The same instability occurs if one attempts an independent, element-wise extrapolation of the baseline coefficients to the CBS limit. Second, the coefficients from Sec.~\ref{subsec:cauchy} include relativistic effects that are absent from the baseline set. This introduces an incompatibility, as the accurate and approximate Cauchy coefficients represent Taylor expansions of two distinct polarizability functions---one including relativistic corrections and the other neglecting them. To circumvent this issue, and to preserve the precise ratios between consecutive terms required for Pad\'e stability, only the single leading Cauchy coefficient ($\alpha_l^{(0)}$) from the baseline set is replaced by the accurate value, while the remaining terms are scaled by the ratio of the accurate $\alpha_l^{(0)}$ to its baseline counterpart. This scaling procedure successfully eliminates singularities, ensuring that all resulting approximants across all considered degrees ($K \le 9$) and polarizabilities ($l=1,2,3$) are smooth functions of the imaginary frequency.

The dispersion coefficients obtained using $\tilde{\alpha}_l^K(i\omega)$ for a given $K$, denoted as $\tilde{C}_6^{K}$, $\tilde{C}_8^{K}$, and $\tilde{C}_{10}^{K}$, are presented in Table~\ref{tab:Cn-convergence}. The convergence with respect to $K$ is rapid, and for $K\ge3$, the individual $\tilde{C}_n^K$ values stabilize to within minor fluctuations of less than 0.1\%. Based on this stable behavior, final recommended values of the dispersion coefficients for two interacting nitrogen atoms, reported in Table~\ref{tab:Cn-error}, are evaluated by taking the arithmetic mean of the results obtained across this specific range of $K$. The magnitude of the observed fluctuations, measured as the standard deviation of the data within the same range of $K$, is included in the uncertainty budget. Another source of error is the uncertainty in the parameters $\alpha_1^{(0)}$, $\alpha_2^{(0)}$, $\alpha_3^{(0)}$, $\zeta_2^{(2)}$, and $\zeta_3^{(2)}$ used to construct Pad\'e approximants for the polarizabilities at imaginary frequencies. The contribution from each parameter is measured as the maximum absolute difference between the recommended value of $C_n$ and the values obtained by shifting that parameter by plus or minus its uncertainty while keeping all other parameters fixed. Assuming that all sources of error are statistically independent, the total uncertainty estimate is obtained by taking the square root of the sum of squares of the individual contributions. An analysis of the uncertainty budget presented in Table~\ref{tab:Cn-error} reveals that errors in the dispersion coefficients are heavily dominated by the uncertainties in the static polarizabilities $\alpha_l^{(0)}$. Notably, these contributions significantly outweigh the numerical noise arising from the use of Pad\'e approximants of various degrees during the Casimir--Polder integration, leaving room for further improvement by increasing the accuracy of $\alpha_{l}^{(0)}$. In contrast, the results are remarkably insensitive to errors in the asymptotic constants $\zeta_l^{(2)}$, whose contributions are at least two orders of magnitude smaller than the effects of $\alpha_l^{(0)}$ and are thus entirely negligible.

\begin{table}
\caption{\label{tab:Cn-convergence}
Dispersion coefficients $\tilde{C}_n^K$ for two nitrogen atoms calculated using $2^l$-pole dynamic polarizabilities $\tilde{\alpha}_l^K(i\omega)$ represented as Pad\'e approximants of degree $K$, see Eq.~(\ref{pade-spec}).
}
\begin{ruledtabular}
\begin{tabular}{cd{4}d{3}d{1}}
$K$ & 
\multicolumn{1}{c}{$\tilde{C}_6^K$} &
\multicolumn{1}{c}{$\tilde{C}_8^K$} &
\multicolumn{1}{c}{$\tilde{C}_{10}^K$} \\
\hline\\[-1.2em]
1 & 24.1379 & 508.812 & 13501.5 \\
2 & 23.9835 & 513.048 & 13674.3 \\
3 & 23.9658 & 514.886 & 13730.4 \\
4 & 23.9614 & 514.932 & 13728.7 \\
5 & 23.9285 & 514.825 & 13728.5 \\
6 & 23.9698 & 515.380 & 13737.4 \\
7 & 23.9619 & 515.300 & 13736.1 \\
8 & 23.9537 & 515.203 & 13734.6 \\
9 & 23.9596 & 515.277 & 13735.8 \\
\end{tabular}
\end{ruledtabular}
\end{table}

\begin{table}
\caption{\label{tab:Cn-error}
Recommended values and uncertainty budget for the dispersion coefficients of two nitrogen atoms. The ``mean($K\ge3$)'' and ``std($K\ge3$)'' rows present the mean value and standard deviation of the data from Table~\ref{tab:Cn-convergence} for $K\ge3$. Subsequent rows ($\Delta\alpha_l^{(0)}$, $\Delta\zeta_l^{(2)}$) list deviations from the recommended values of $C_n$ due to uncertainties in the corresponding quantities. Numbers in the parentheses in the last row denote the uncertainty estimates at the last digit(s).
}
\begin{ruledtabular}
\begin{tabular}{cd{7}d{5}d{4}}
contribution & 
\multicolumn{1}{c}{$C_6$} &
\multicolumn{1}{c}{$C_8$} &
\multicolumn{1}{c}{$C_{10}$} \\
\hline\\[-1.2em]
mean($K\ge3$)          & 23.9558    & 515.153   & 13733.5    \\
std($K\ge3$)           &  0.014     &   0.22    &     3.9    \\
$\Delta\alpha_1^{(0)}$ &  0.049     &   0.52    &     8.7    \\
$\Delta\alpha_2^{(0)}$ &            &   2.7     &    55.     \\
$\Delta\alpha_3^{(0)}$ &            &           &    47.     \\
$\Delta\zeta_2^{(2)}$  &            &   0.003   &     0.1    \\
$\Delta\zeta_3^{(2)}$  &            &           &     0.1    \\
\hline\\[-1.2em]
total                  & 23.956(51) & 515.2(28) & 13733.(73) \\
\end{tabular}
\end{ruledtabular}
\end{table}

\subsection{Comparison with the literature}
\label{subsec:literature}

\begin{table*}
\caption{Comparison of the dispersion coefficients $C_6$, $C_8$, and $C_{10}$ with the available literature data. A brief overview of the theoretical method used in the previous papers is given below the table. Our results are shown in the last row.
\label{tab:lit_comp}}
\begin{ruledtabular}
\begin{tabular}{lccc}
& \multicolumn{1}{c}{$C_6$} & \multicolumn{1}{c}{$C_8$} & \multicolumn{1}{c}{$C_{10}$} \\
\hline
Krauss and Neumann$^a$ (1976) \cite{krauss19765sigma} & 32 & 631 & 16253 \\
Zeiss and Meath$^b$ (1977) \cite{zeiss1977dispersion} & 24.12 & & \\
Margoliash and Meath$^b$ (1978) \cite{margoliash1978pseudospectral} & 24.10 & & \\
Partridge et al.$^c$ (1986) \cite{partridge1986theoretical} & 24.12 &  475.5  &12247.9 \\
Fowler et al.$^d$ (1990) \cite{fowler1990c} & 24.00 &  &  \\
\multirow{2}{*}{Hettema and Wormer$^e$ (1990) \cite{hettema1990frequency}} & 23.944 &  478.48 & \\
 & 24.033 &  475.01 & \\
Chu and Dalgarno$^f$ (2004) \cite{chu2004linear} & 24.2 & & \\
\multirow{2}{*}{Tscherbul et al.$^g$ (2010) \cite{tscherbul2010collisional}} & 24.0 & & \\
 & 23.36 &  &  \\
Gould and Bu\v{c}ko$^h$ (2016) \cite{gould2016c} & 25.7 & & \\
\hline
this work & 23.956(51) & 515.2(28) & 13733.(73) \\
\end{tabular}  
\end{ruledtabular}
\vspace{-0.3cm}
\begin{flushleft}
$^a$ Based on an approximate relationship between $\alpha_0$ and $C_6$; $C_8$ and $C_{10}$ estimated using the method of Ref.~\onlinecite{starkschall72} \\
$^b$ Semi-empirical dipole oscillator strength distributions (DOSD) approach \\
$^c$ The coefficient $C_6$ taken from Ref.~\onlinecite{zeiss1977dispersion} and $C_8$, $C_{10}$ are scaled results from Ref.~\onlinecite{krauss19765sigma} \\
$^d$ Multiconfigurational self-consistent field method with $1s^2[2s2p3s3p3d]^5$ active space and $8s6p5d2f$ atomic orbital basis \\
$^e$ Time-dependendent coupled Hartree--Fock method with two different orbital basis sets including up to $f$-type functions \\
$^f$ Time-dependent density functional theory (OEP-SIC functional) with empirical correction factor \\
$^g$ From fitting of the interaction potential for the $^7\Sigma_u^+$ state, two different methods for calculation of interaction energies \\
$^h$ Time-dependent density functional theory with frequency rescalling
\end{flushleft}
\end{table*}

In Table~\ref{tab:lit_comp} we compare the dispersion coefficients obtained in this work with the data available in the literature and provide a brief description of the methods used in the cited papers.

The first dispersion coefficient $C_6$ is the simplest to evaluate among all $C_n$, and it has been the subject of numerous previous studies. The data in the literature cluster within the interval $24.0-24.2$, with only a few exceptions, depending on the method used. Our result $C_6=23.956(51)$ is slightly smaller than this, but a more direct comparison would require taking into account the uncertainties of previous calculations, which are usually not reported. Only Zeiss and Meath \cite{zeiss1977dispersion} and Margoliash and Meath, \cite{margoliash1978pseudospectral} two works based on the dipole oscillator strength distributions (DOSD) approach, stated that their accuracy is no worse than 1\%. If this is taken into account, our result is fully consistent with their calculations. Additionally, Zeiss and Meath \cite{zeiss1977dispersion} report a bound, $24.9\pm 1.7$, for the exact value of $C_6$ based on analysis of Pad\'e approximants used in the DOSD approach. This bound is also consistent with the results of the present work. Overall, our value agrees best with calculations of Hettema and Wormer \cite{hettema1990frequency} based on the time-dependent coupled Hartree--Fock method. Their results differ from ours by less than 0.1\% or about 0.3\%, depending on the basis set used in Ref.~\onlinecite{hettema1990frequency}. However, it is worth pointing out that the method used in Ref.~\onlinecite{hettema1990frequency} neglects electron correlation effects in the reference state, so such an astonishingly good agreement might be in part due to accidental error cancellation.

The results for higher dispersion coefficients, $C_8$ and $C_{10}$, are much less numerous in the literature. Moreover, some of them are not direct calculations, but rather estimates based on approximate relationships between $C_n$ and other quantities which are easier to compute, see, for example, Krauss and Neumann. \cite{krauss19765sigma} In the case of the $C_8$ coefficient, we are aware of only a single direct calculation by Hettema and Wormer. \cite{hettema1990frequency} Their results are smaller by roughly $7-8\%$ than our data. In the case of the $C_{10}$ coefficient, we were not able to find any paper in which this quantity was directly calculated. The estimated values of Krauss and Neumann \cite{krauss19765sigma} and Partridge et al. \cite{partridge1986theoretical} are based on scaling using lower coefficients or approximate relationships. They deviate from our result for $C_{10}$ by $10-20\%$.

\section{Conclusions}
\label{sec:conclusions}

In this work, we have performed first-principles calculations of dynamic polarizabilities of the nitrogen atom and long-range dispersion coefficients, which describe the interaction of two nitrogen atoms at large internuclear distances. To achieve the stringent accuracy demanded by metrological applications, we developed customized Gaussian basis sets, denoted as a$X$Z(pol) and da$X$Z(pol). By actively optimizing the augmenting diffuse functions with respect to a Hylleraas-like functional, we achieved a highly stable and systematic convergence of dipole, quadrupole, and octupole polarizabilities without the prohibitive computational costs of standard heavily-augmented basis sets. The frequency-dependent polarizabilities were expanded into Cauchy series, with the constituent Cauchy coefficients $\alpha_l^{(2k)}$ calculated through a rigorous composite scheme. This scheme captured dominant valence correlation effects using the FCI method, while systematically incorporating core-valence correlation effects up to the ae-FCI level and scalar relativistic corrections via the DKH2 method. By directly evaluating the response equations at the FCI level, we bypassed the numerical instabilities associated with polynomial fitting of higher-order Cauchy coefficients. The resulting coefficients, particularly for the dipole polarizability, exhibit tight relative uncertainties on the order of $0.2\%$ or better. Furthermore, we determined the $C_6$, $C_8$, and $C_{10}$ dispersion coefficients using the Casimir--Polder formalism. The requisite polarizabilities at imaginary frequencies were constructed using Pad\'e approximants. We demonstrated that anchoring these approximants with highly accurate static polarizabilities significantly accelerates convergence toward the exact limits, yielding highly reliable dispersion coefficients with robust error budgets. In subsequent papers of this series, we will focus on the properties of the nitrogen molecule, and the data provided here will be crucial, for example, for modeling the long-range tail of the N$_2$ interaction potential.

\section*{Supplementary Material}

See the Supplementary Material for the detailed composition of the a$X$Z(pol) and da$X$Z(pol) basis sets, the list of frequencies used for fitting the Cauchy coefficients, the raw results and a summary of the calculations of Cauchy coefficients reported in this work, the baseline Cauchy coefficients used to construct Pad\'e approximants, the calculations of asymptotic coefficients $\zeta_l^{(2)}$ describing the behavior of dynamic polarizabilities in the high-frequency limit, and the derivation of the analytic formula for the $\zeta_l^{(4)}$ coefficients.

\begin{acknowledgments}
The project (22IEM04 MQB-Pascal) has received funding from the European Partnership on Metrology, co-financed from the European Union’s Horizon Europe Research and Innovation Programme and by the Participating States. We gratefully acknowledge Poland's high-performance Infrastructure PLGrid (HPC Centers: ACK Cyfronet AGH, PCSS, CI TASK, WCSS) for providing computer facilities and support within computational grant PLG/2025/018692. The authors also thank Pozna\'n Supercomputing and Networking Center for the computational grant pl0458-01. 
\end{acknowledgments}

\section*{Data Availability Statement}

The data that support the findings of this study are available within the article and its supplementary material, and openly available in Zenodo repository at \url{https://doi.org/10.5281/zenodo.22662830}, reference number 22662830.

\appendix

\section{Direct evaluation of the Cauchy coefficients}
\label{sec:appb}

Direct calculation of the Cauchy coefficients at the FCI level of theory is based on the expansion of the $2^l$-pole polarizability in powers of $\omega$. Starting with the definition given in Eq.~(\ref{alphaw2}) we expand the resolvent operators according to the formula
\begin{align}
    \big( H - E_0 \pm \omega \big)^{-1} = \sum_{k=0}^\infty 
    \frac{(\mp\omega)^k}{\big( H - E_0 \big)^{k+1}},
\end{align}
which is convergent provided that $\omega$ is smaller than the first resonance frequency of the system. Inserting this expression into the definition of the polarizability, Eq.~(\ref{alphaw2}), and collecting terms multiplying the same power of $\omega$, one sees that the odd powers of $\omega$ vanish. We are left with:
\begin{align}
\begin{split}
    \alpha_l(\omega) &= 2\sum_{k=0}^\infty 
    \langle0|Q_{l0}\,\frac{\omega^{2k}}{\big( H - E_0 \big)^{2k+1}}\,Q_{l0} |0\rangle,
\end{split}
\end{align}
from which we recognize the Cauchy coefficients as
\begin{align}
    \alpha_l^{(2k)} = 
    2\langle0|Q_{l0}\,\big( H - E_0 \big)^{-(2k+1)}Q_{l0} |0\rangle.
\end{align}
Direct evaluation of this formula using the sum over states would require complete diagonalization of the FCI Hamiltonian, which is not feasible in practice. To avoid this problem, we introduce a family of response functions, denoted by the symbol $|r_k\rangle$ where $k\geq 0$ is the order, defined recursively as
\begin{align}
\label{r0}
    |r_0\rangle &= Q_{l0} |0\rangle, \\
\label{rk}
    \big( H - E_0 \big)|r_{k+1}\rangle & = |r_{k}\rangle.
\end{align}
Using this definition of the response functions, the Cauchy coefficients take a simple form
\begin{align}
    \alpha_l^{(2k)} = 2\langle r_k|r_{k+1}\rangle.
\end{align}
In our calculations, both $|0\rangle$ and $|r_k\rangle$ are represented by the FCI expansion, i.e., a linear combination of all $N$-electron Slater determinants that can be assembled within a given orbital basis set. As a result, Eq.~(\ref{rk}) is represented as a set of linear equations that can be solved using standard methods, yielding the expansion coefficients without complete diagonalization of the Hamiltonian. Therefore, the calculation of all Cauchy coefficients $\alpha_l^{(2k)}$ up to a given $k$ requires finding the electronic ground-state wavefunction $|0\rangle$, followed by solving $k+1$ sets of response equations.

\bibliography{nitrogen-paper1}

\end{document}